# ESTIMATING THE FDI IMPACT ON ECONOMIC GROWTH AND EXPORT PERFORMANCES OF THE EUROPEAN ECONOMIES IN TRANSITION

**Assist. Olivera Kostoska Ph.D Student**
University „St. Kliment Ohridski"
Faculty of Economics
Prilep, Republic of Macedonia
**Assoc. Prof. Pece Mitrevski Ph.D**
University „St. Kliment Ohridski"
Faculty of Technical Sciences
Bitola, Republic of Macedonia

**Abstract:** Within the last two decades, Foreign Direct Investment (FDI) has been observed as one of the prime instruments in the process of restructuring the European economies in transition. Many scholars argue that FDI is expected to be a source of valuable technology transfer thus might certainly have positive effects on host country development efforts. Nonetheless, there are no clear-cut findings about the FDI genuine performances in supporting the economic growth, productivity and export improvements within the European transition countries. Using a large and comprehensive data set, we will therefore analyze the linkage between FDI and above mentioned variables, so as to recommend national policy appropriate measures aimed at averting negative and strengthening the positive FDI spillovers.

## Introduction

Foreign capital is perceived to be a fundamental source of financing the necessary economic reforms within the European transition countries. Thus, alongside with the possible increase or substitution the domestic savings, FDI might have a great impact on the external financial position. Put differently, early FDI privatization revenues have improved the net foreign reserves in many transition countries (Hungary and Czech Republic at most), additionally to their external debt cutback. At the same time, FDI has facilitated financing the current account deficits since seems to be relatively stable, export supportive and non-debt creative. On the other hand, the host economy could attain positive gains from the interests of the country investor to protect its own capital against all the external threats or western protectionism (Stankovsky, 1995). Equally important to the macroeconomic function, FDI is also perceived to greatly improve the competitiveness at a micro level by proceeding the organizational and technology transfer. At this point, FDI performs completely different effects in comparison to portfolio investment or direct lending since looks as if being less reversible, thus facilitating the economic growth and prosperity of the countries (Sinn and Weichenrieder, 1997).

Many scholars, notwithstanding, argue that FDI positive spillovers are not to be always straightforward, but few or far between. Some recent findings suggest that a number of foreign investment might result into negative spillovers if induces the host enterprises to close down since those preserved not to finance the technology upgrade. On the other hand, FDI might not always help financing the balance of payments particularly if capital flows come into non-tradable sector. Put differently, FDI could impose nominal and real exchange rate appreciation and therefore diminish the export competitiveness.

Hence, the main purpose of this paper is to assess the consequences of investment liberalization for the European transition economies. Allowing for the competitiveness as an "ability to grow in an open setting", we will also address some of the main straits which FDI has been augmenting through the growth and export performance of the countries. After the short outline about theoretical and empirical findings on the particular matter, we consider to present an overview on recent trends and factors attracting FDI within Section 2. The statistical correlation between FDI and economic growth is to be an area under discussion in Section 3, while Secton 4 will look at FDI sectoral composition and related implications on export competitiveness. Section 5 exploits the long-term FDI effects, as well as the national measures a country has to undertake in order to improve the positive and diminish the negative FDI spillovers. Finally, we exhibit the short conclusions and recommend some national policy improvements recounted to the respective issues.

**FDI, economic growth and export performances: theoretical and empirical considerations**

The standard theoretical Solow model (1957) indicates that economic growth is dependent upon the growth rates of capital and labor weighed by their shares of income in addition to the level of technology progress.

$$\frac{\Delta y}{y} = \theta * \frac{\Delta K}{K} + (1 - \theta) * \frac{\Delta N}{N} + \frac{\Delta A}{A}$$

where *Y* stands for GDP, $\theta$ and $(1 - \theta)$ symbolize capital and labor share of income respectively, *K* indicates the amount of capital, *N* is equal to labor force and *A* denotes the level of technology. Additionally, Solow found that output growth has outpaced the weighted average increase in capital and labor inputs. Thus, if the output growth rates, labor force and capital stock are well identified, the growth of technical progress may possibly be estimated, too. On the other hand, technical progress depends upon the R&D expenditures, as well as the technological improvements with FDI as their best source of transmission.

The theory of endogenous growth, however, argues that FDI has an effect on economic growth passing either through the variable such as R&D or the human capital education (Romer, 1986). In consequence, the technology transfer might get moving the intermediate products development, increase the quality, make the international research cooperation possible, as well as establish some new forms of human capital (Aghion and Howitt, 1992).

Although most of those representatives prove the positive correlation between the technology transfer and economic growth, very few of them really assess the genuine role of FDI in creating the particular spillovers (Mello, 1997). [14] Namely, some empirical studies based upon the endogenous growth theory indicate that FDI may possibly allow the technology transfer and improve the economic growth if only the host economy has a minimum threshold of human capital (Borensztein, De Gregorio, Lee, 1998). On the other hand, the latest findings suggest that host R&D efforts have a better impact on productivity growth than foreign, so the other transmitters of technology transfer have to be observed more closely (Keller, 2000). At the outset, some effects of scale uncovered within the industry data point toward superior meaning of direct technology transfer than the spillovers, but the lack of statistically significant proofs indicate that conduits of technology transfer are not to be comparable but rather supplementary one to each other.

### Recent trends and factors attracting FDI flows into European transition economies

There is widespread evidence that foreign capital flows are usually dependent upon factors differing among the various countries. Thus, FDI might be attracted by the terms of demand, political or macroeconomic stability, market size and liberalization, the factor endowment, skilled labor force, privatization, investment risk etc. FDI flows into European transition economies have been insignificant until the early nineties. The extent of reforms undertaken afterwards, as well as polices designed to engender an investment friendly environment are to be the major drivers of FDI flows into these countries (Figure 1). Consequently, the group of transition economies considered as EU candidates (Czech Republic, Hungary and Poland) accounted for 60% of all FDI inflows, principally drawn by the privatization process of state owned enterprises.[15]

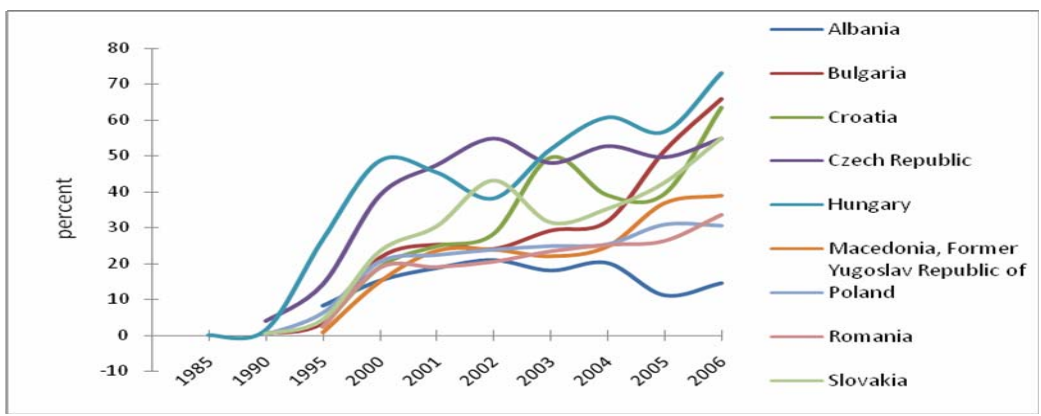

**Figure no. 1 FDI inward stocks as a percentage of GDP**

The other advantage for these countries is the geographical propensity to the Western European markets, but for the most part free trade provisions of the EU Association Agreements. FDI flows into the other EU transition economies, however, fell far from belief until the 1999 when succeeded to receive annual investment in approximate 4% of GDP. The major objective of the national policy makers was attracting the large scale foreign capital aimed at achieving faster economic growth and better integration into the world economy. The subsequent EU accession announcement seemed to be the foremost rationale for a large amount foreign investment into some transition countries such as Romania and Bulgaria. Consequently, the total FDI inflows kept getting higher despite the financial crisis during 1997-1998 expressing the long-run strategic prospects of foreign investors, as well as the opportunities put across the depressed asset prices. Distinctive point in this process was the Kosovo crisis, although the major privatization efforts did go forward. Noteworthy is to mention that 70% of these projects have been accomplished within the industry (electronics, food industry and raw materials extraction) and 30% in the service sector (finance, insurance and telecommunications). In addition, the utmost share of manufacturing noticed Romania (78.1%), followed by Poland (63.3%) and Czech Republic with the highest share of car industry, chemicals and food processing. On the other hand, data available for service sector confirmed that investment in finance took the major part in Poland and Slovak Republic (19% in 1995), as well as Hungary and Czech Republic (11.2% and 7.6%, respectively). At the same time telecommunication sector accounted for 12.1% of all the FDI flows within the European economies in transition, with special emphasis placed on Hungary and Czech Republic which have accounted for 90% of all the foreign capital in the particular sector.

---

[14] For further details see also Coe, D. and Helpman, E., "International R&D spillovers", *European Economic Review,* Vol. 39, pp. 859-887, 1995.

[15] Hunya, G.,*"International Competitiveness : Impacts of FDI in CEECs ",* The Vienna Institute for International Economic Studies (WIIW), Research Reports, No. 268, Vienna, 2000.

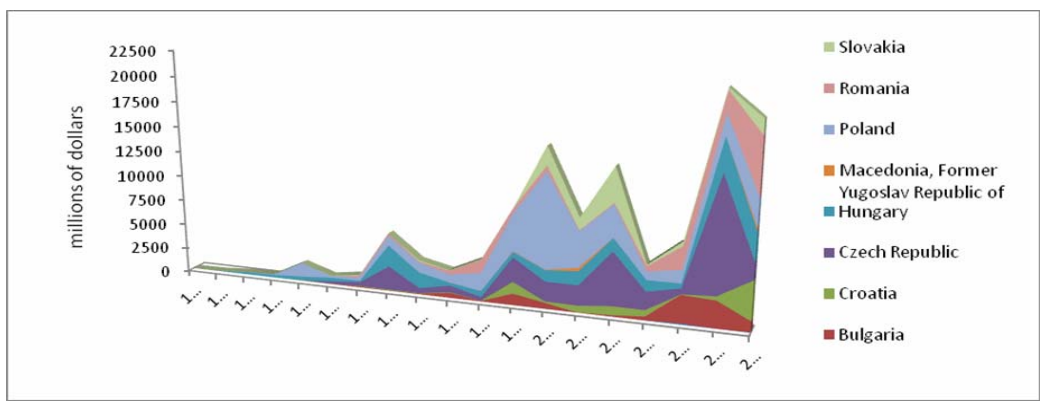

**Figure no. 2 Cross-border M&A purchases in European transition economies (1988-2006)**

Some of these countries have also attracted foreign investment in trade and tourism such as Bulgaria (19.5%) and Slovak Republic (19.5%). Finally, worth mentioning is that cross-border mergers and acquisitions (M&A) have been the principal approach for the majority of FDI entry into these privatization projects, predominantly caused by the extensive wave of liberalization and process of deregulation. The large-scale decline of the stock value traded on the world markets implied the M&A fall by 38% in 2002, in addition to twisting down of privatization within some transition economies (Figure 2). During the particular period, Greenfield investment managed to preserve the total FDI increase mostly in the form of reinvested profits.

**FDI, economic growth and capital accumulation**

Theoretical and empirical findings explained above indicate that countries are likely to achieve higher economic growth if only they are opened to new technologies. FDIs preserve to be one of the best channels to technology transfer and therefore is supposed to have an impact either on GDP or gross fixed capital formation (GFCF). Within the last decade opening, European economies in transition have undergone a huge transformation of their systems, followed by structural adjustment, loss of the markets, as well as the economic performances turn down. These problems have been resolved to some extent in the middle of decade when Poland, Slovak Republic, Romania and Czech Republic noticed the first more dynamic growth rates. After flourishing implementation of the so called "austerity package", Hungary has also achieved a 2% growth rate, as a starting point to further economic upswing. Generally spiking, almost all the European economies in transition succeeded to stop marking negative growth rates since 1995, which have marked the end of the transitional recession within the region (Figure 3). Put differently, the average growth rates of the European transition economies surpassed the world's average in about 1%, so every single excluding Croatia attained a full membership intro the European Union.[16]

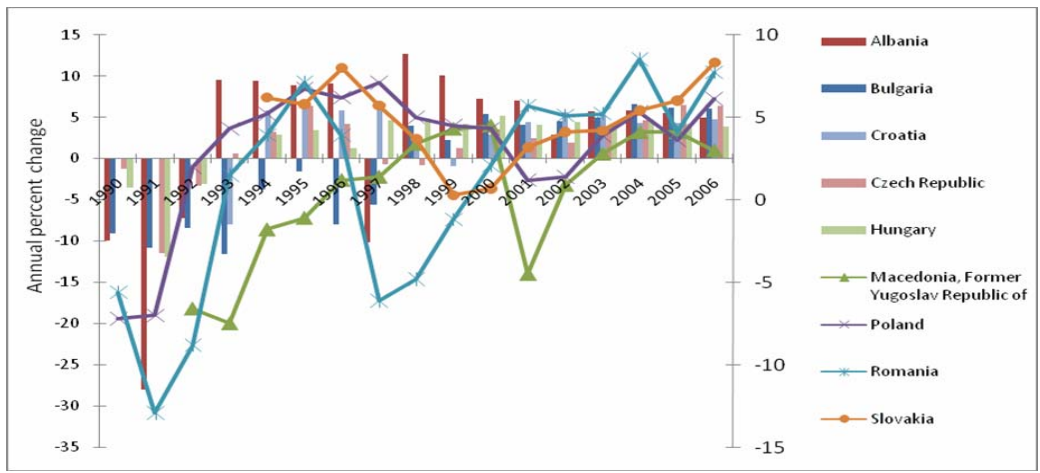

**Figure no. 3 Annual GDP percentage change of the European transition economies**

Looking at the data analysis (Figure 4), one can find positive but slightly significant correlation between FDI and economic growth for the selected European economies in transition. The reason behind come across the period taken into study which is distinguished by the level of stabilization polices assumed, as well as the strength of import demand within the key trading partners. Some transition economies, however, have improved their economic performances pursuing the FDI driven export growth. In fact, exports have been considered the most dynamic component of the final demand extremely going beyond the united contribution of investment and consumption.[17] Consequently, many countries have experienced a positive correlation between GDP and export growth, particularly if foreign companies affect a large share of exports. Empirical findings about the positive correlation between FDI and long-term growth put forward the possibility for the very same achievement within the transition economies, as well.

---

[16] The reasons either for Croatia or other transition economies failure not to access the European Union are to be found in some political or structural reforms instability emerged principally by the recurrent disequilibrium between gross domestic savings and investment.

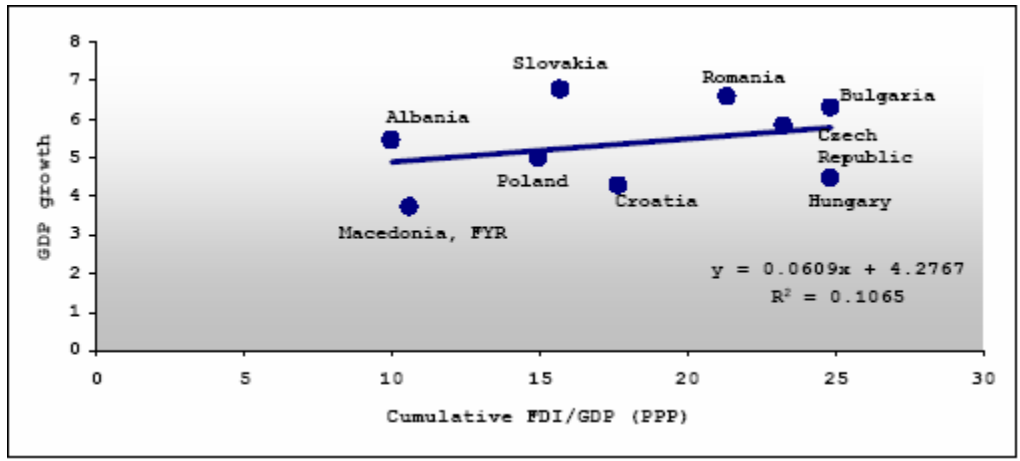

**Figure no. 4 GDP growth and ratio of cumulative FDI inflows to GDP (PPP)[18]**

Thus, FDI/GDP ratio of Eastern Europe has increased from 0% to 4% during the period 1997-1999 (in nominal GDP terms). In addition, elasticity estimated by Borenzstein suggests an increase of some 0.7-1.3 percentage points in the long-term GDP/capita growth, especially in Czech Republic and Hungary. The particular elasticity reveals the average human capital in developing countries. In other words, FDI is more emphatic in the countries with higher average level of human capital. Transition economies have not been faced with the lack of exacting capital, but had certain problems with the process of reforms. Thus, Borenzstein elasticity is not reliable enough to judge for the countries which haven't established long-term market system, such as transition ones. The above mentioned is considered to be the main reason one could not determine a direct linkage between amount of FDI inflows and dynamic economic growth. In other words, countries with high FDI inward stocks have run off lower growth rates (Hungary, Czech Republic), despite those vigorously growing economies (Slovakia and Poland) which have recorded smaller amount of foreign capital (Figure 4). The reason behind emerged from the basic premise that FDI inflows in European transition economies were motivated from the cheap labor force. Nevertheless, the most FDI inflows aimed at gaining market access while the labor costs seemed to be of secondary importance. Additionally, the mode of FDI entry into European transition economies is supposed to be the second reason behind this weak correlation between FDI inflows and economic growth. Thus, M&As whose share in FDI is very high, suggest smaller impact on economic growth, since they correspond to

---

[17] FDI driven export growth have been mostly responsible for the economic improvements in Hungary during the nineties. Czech Republic noticed the very same results even though GDP has been actually constricted owing to domestic absorption collapse.

[18] Average growth of real GDP, 2003-2006 (estimates). FDI inflows are cumulated from 1996 to 2006. GDP (PPP) refers to 2006.

**Business Statistic – Economic Informatics**

a transformation of ownerships, rather than a new capital addition. They will certainly spawn positive spillovers and improve the economic performances of the privatized manufacturing firms save in a longer timeframe.

The importance of FDI might be also observed if one estimates the relation to gross fixed capital formation (predominantly the private corporate investment) and expresses the GFCF amount as a proportion of GDP (Table 1). Thus, private sector investment in Hungary with privatization revenues excluded declined from 29% in 1991 to 17% in 1994 (Hunya, 1995).

**Table no 1 Gross fixed capital formation (% of GDP)**

| GFCF % of GDP | | | | | |
|---|---|---|---|---|---|
| Country | 2002 | 2003 | 2004 | 2005 | 2006 |
| Albania | 32 | 25 | 26 | 24 | 26 |
| Bulgaria | 20 | 22 | 23 | 27 | 32 |
| Croatia | 29 | 31 | 31 | 31 | 30 |
| Czech Republic | 28 | 27 | 27 | 26 | 27 |
| Hungary | 26 | 25 | 26 | 24 | 23 |
| Macedonia, Former Yugoslav Republic of | 21 | 20 | 21 | 20 | 22 |
| Poland | 19 | 19 | 20 | 19 | 20 |
| Romania | 22 | 22 | 22 | 23 | 24 |
| Slovakia | 29 | 25 | 26 | 29 | 29 |

*Source: UNCTAD*

On the other hand, Stankovsky (1995) has estimated the foreign shares on 10% for Poland and Czech Republic in 1992, as well 4% for the Slovak Republic if privatization revenues included. Moreover, in-debt analyses on investment financing

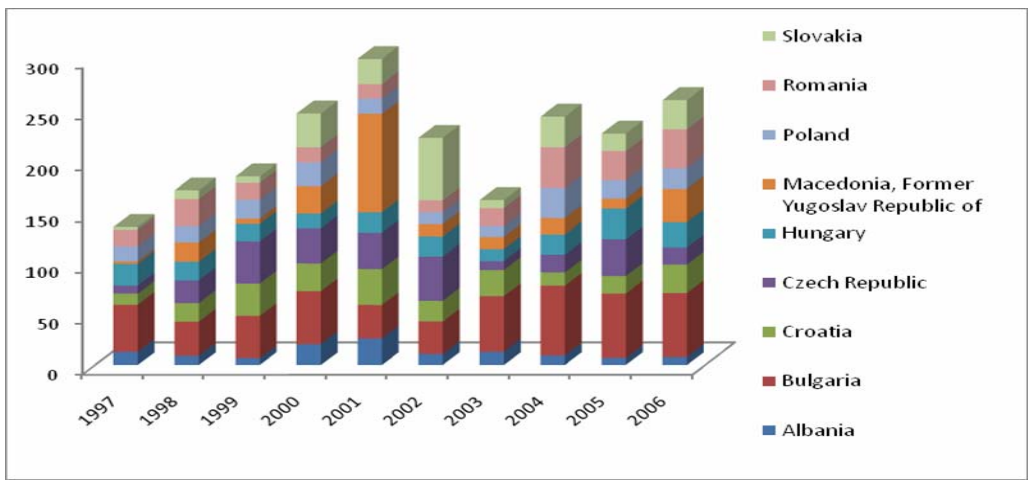

**Figure no 5 FDI inward flows as a percentage of GFCF**

carried out by Quaisser (1995) and Nesvera (1995) indicate that foreign capital did not exceed 10% as a proportion of the real investment in Czech Republic, while FDI in

Poland has started with 11.4% of GFCF in 1993, but come out to fall below 10% in 1994. Albeit many fluctuations of investment rates in European transition economies, evidence suggests that they are much higher compared to those of Latin America, but lower measured up to Asia, particularly if China included. Although European economies in transition have not achieved impressive investment rates, they still seem to be satisfactory in getting convergence with the European developed countries. In principle, transition economies, equally to middle-income developing ones have to attain investment rates in about 25% of GDP if want to pull off a sustainable economic growth (ELAC, 2000 and UNCTAD, 2001). The required investment rates, however, may not be achieved if the country props up the gross domestic savings only, since they are not considered to be as much as satisfactory in private sector. Therefore, European transition economies put much effort to attract foreign capital so as to close up the existing disequilibrium between investment and gross domestic savings, as well as to set off the private and public domestic investment in GFCF (Figure 5). Despite the other sources of finance, FDI might also internalize foreign savings and does not perform any subversive side-effects for the host economy. Additionally, the increased capital stock is believed to have more direct effects on economic growth than the technological progress. [19] Affecting the domestic savings as a supplementary source of finance, FDI might certainly help transition economies to create adequate conditions for getting on the trace of further technological improvements.

### FDI, sectoral decomposition and export performances

National economies may achieve a sustainable long-term growth if only manage to attain persistent improvements in productivity and efficiency. Transition economies place special emphasis on reforms in order to generate higher growth rates, unlike developed ones with intra-industry expansion as the foremost rationale. Yet, intra-industry augmentation must not be undervalued for transition economies as well primarily due to differentials among the sectors for possible technical progress and relative productivity growth. FDI is the most convenient form of capital inflow that may perhaps add to productivity either by management or technology transfer and therefore increase the possibility to dislodge the current account disequilibrium especially if export-oriented. Thus, FDI has helped financing 86% of the fourfold increase in current account deficits of the European transition economies during the nineties.[20] This points toward the quite favorable disposition of foreign investment since looks as if being relatively stable, export promoting and non-debt creating. These commonly positive FDI aspects usually related to more dynamic export growth may also contribute to getting better perception about the host country creditworthiness. The potential benefits of FDI should not be overestimated, however, since the rising tendency of exports within the European transition economies was partially imposed by the upswing in Western Europe in 1994, but also the domestic (supply) factors. In other words, private consumption, stimulated by the real wages raise, has played an important role, as well. Furthermore, FDI may possibly impose the same risks such as the other capital flows. Put differently, foreign inventors might indispose their capital into non-tradable sectors and dependant on the exchange rate regime to create nominal and real exchange rate appreciation so as to weaken the export competitiveness. From this point of view, European economies in transition have to set down the inflation rates not to be much different from the world's average if want to preserve the competitiveness in terms of prices, principally for the sectors with the largest share in their exports. The FDI flows into service sector might be particularly important if one considers the level of transition economies development. The bulk of foreign investment in telecommunication, banking and different business services is supposed to spawn positive spillovers and enhance the exports. Thus, FDI is not promptly assessable in this sector as regards the productivity gains, nevertheless, might possibly add to improving the efficiency of the wider business climate. Put differently, developments of information and communication technology possibly will generate positive spillovers in all the other sectors and promptly change the tradability of information-related services. In addition, improvements in services of physical and technological infrastructure, as well as the local-bound tourism may possibly be an important resource of revenue.

---

[19] For more details see also Eichengreen, B., *"Capital flows and crises",* The MIT Press, 2004.

[20] For several years Poland has turned to relative equilibrium noticing current account deficit in about 1.5% of GDP. Czech Republic and Slovakia have also kept the deficits under control with relative figures equal to 2.5% and 6%, respectively. The last reformers, Bulgaria and Romania have recently marked the deficits cut-back although they seemed to be very far from equilibrium.

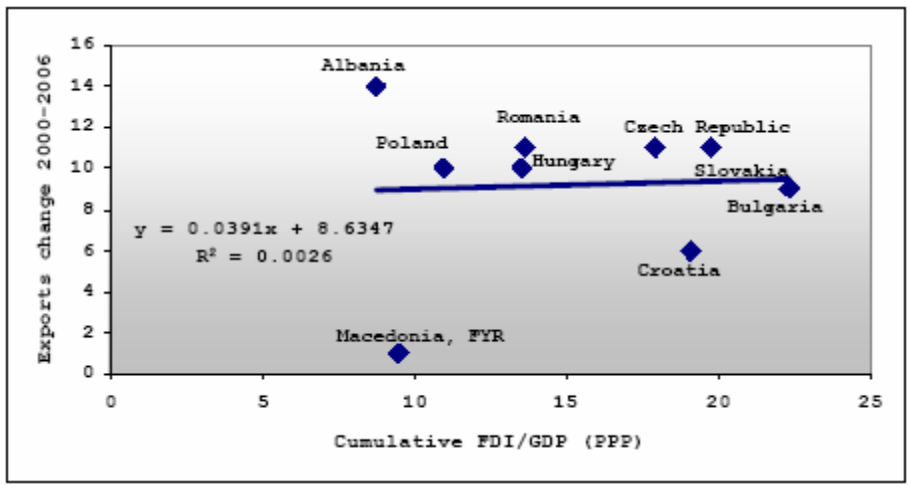

**Figure no 6 Exports growth and ratio of cumulative FDI inflows to GDP (PPP)[21]**

      The recovery of industrial output, as well as the trade liberalization in many European transition economies raised a large amount of FDI flows into the manufacturing sector.[22] The principal motives are to be found within the proximity of the larger European market, the highly skilled labor force, but most of all the labor costs per production unit (ULC) which are perceived to be the most advantageous in West European comparison. The empirical findings suggest that countries which have traced more FDI in manufacturing have recorded an improvement in export competitiveness weighed against those with FDI flows in services (Figure 6). Thus, 70% of manufacturing sales in Hungary were credited to foreign investment in 1998, so the country tripled the exports during the period of subsequent five years. These figures actually revealed the authorities' privatization strategy indisposed towards foreign investors appeal as potential bidders using various forms of concessions such as tax and

---

[21] Exports change in 2006 relative to 2000. FDI inflows are cumulated from 2000 to 2006. GDP (PPP) refers to 2006.
[22] World Investment Report 1997

other forms of holidays. Yet, foreign owned firms have achieved the majority of total factor productivity gains producing time and again negative spillovers for the domestic owned enterprises (Aitken and Harrison, 1999). The reason behind are to be found in the technological gap between the two groups of firms and lack of possibility for the domestic ones to obtain necessary financing for upgrading their equipment and managerial know-how. Over the time as competition increases, backward and forward linkages started to grow up and outsourcing turn out to be more prevalent. Noteworthy is to mention that productivity lag of the countries might also appear on account of different relative prices of intermediate goods between foreign and domestic owned enterprises allowing for the transfer pricing. Yet, productivity gap is likely to be less significant if measured by value added per employee. Thus, labor productivity of local firms was estimated to two thirds of the foreign in Czech Republic and 80% of those in Poland at some stage in 1993.

The sectoral FDI composition in Czech Republic has been evenly scattered between services and manufacturing, thus the country has less frenetic growth in exports relative to FDI. Similarly, Croatia faced smaller effects on exports competitiveness, although the country experienced large FDI inflows. Those, however, have been mostly of the local market seeking type, such as retail and financial intermediation, with no large perspectives to change the export structure immediately. At the same time manufacturing comprises just one third of FDI in Bulgaria.

**Table no 2 R&D contents of exports**

| Country | 2002 | 2003 | 2004 | 2005 |
|---|---|---|---|---|
| Albania | 1 | 1 | 1 | 1 |
| Bulgaria | 4 | 4 | 4 | 5 |
| Croatia | 12 | 12 | 13 | 12 |
| Czech Republic | 13 | 13 | 13 | 13 |
| Hungary | 25 | 26 | 29 | 25 |
| Macedonia, Former Yugoslav Republic of | 1 | 1 | 1 | 1 |
| Poland | 3 | 3 | 3 | 4 |
| Romania | 3 | 4 | 3 | 3 |
| Slovakia | 3 | 4 | 5 | 7 |

*Source: World Development Indicators*

At the outset, noteworthy is to conclude that international competitiveness of the European transition economies in average has been greater than before opening to FDI. Thus, foreign owned firms have participated within the Hungarian exports in almost 90%. More intriguing, however, is the lower productivity improvements of the domestic in comparison to foreign owned sector, as well as the sort of the so called "postponed spillover effects syndrome". The technological gap running down is seemed to be the best way to improve the competitiveness of the local enterprises, as well as to make the growth and convergence sustainable. An additional way is following the possibility to alter the capital intensity of the exports (Table 2). Thus, the premature reformers such as Poland, Hungary and Czech Republic have granted the highest technology transfer while receiving the most FDI in manufacturing. They have increased the R&D composition of their exports, even though stayed within the low value added section of the high technology sectors. Time and again, the lack of FDI manufacturing inflows in Croatia was effectuated in no technology transfers and less changes in value added composition of Croatian exports.

### National policy measures and Foreign Direct Investment

FDI driven productivity, exports and growth expansion stipulate many transition economies to implement various measures aimed at attracting foreign capital. Those include tax incentives or perk-filled economic zones, upgrading the overall business environment, upholding the predictable and transparent rules, starting the export processing zones etc. Evidence suggests that many of the above noted measures, however, might be inappropriate and less suited for the particular necessities. Thus, some of the tax incentives preserve to reduce regional disparities and increase spending on education and R&D. Yet, erosion imposed within the tax bases may augment the tax burden for the others, increase corruption and support the potentially unprofitable activities. The effects of special economic zones are mixed in some way as regards the experience of different countries. In other words, the advantages in terms of new employments or linkages with the local firms might be limited as the zones are likely to create only highly skilled jobs. Taking into consideration the above limitations the best way to attract investors is to improve the business environment as a whole principally throughout the predictable polices, transparent legal system and simple licensing regime. At the same time, domestic firms have to be also supported so as to compete more effectively with the foreign ones (escape negative spillovers, such as bankruptcy or become more dynamic partners). These activities are indispensable to be applied since FDI in parallel with the EU accession driven reforms enabled many transition economies to become fully-member states of the European Union.

### Conclusions and policy recommendations

The evidence in the paper suggests that many European economies in transition have attracted significant FDI inflows, but there is also a rising disproportion among the countries within the region. Thus, the low-income transition economies have lagged behind the early reformers (Czech Republic, Hungary and Poland) which have received almost 60% of the total FDI inflows. Statistical analyses performed in the paper have also shown a positive direct FDI impact on growth and export performances particularly those which have received large FDI amounts. Transition economies where FDI are supposed to record a great influence, nevertheless, GDP growth rates are still not satisfactory to promptly shrink the income gaps with some of the EU 15 countries. The empirical findings for developing economies, however, suggest that FDI has a long-run impact on growth, so the very same might be expected for the transition countries, as well. Despite the belief that FDI, among the others, has a great importance in producing positive externalities, analyses presented here suggest that there have been few or no assenting productivity spillovers. In other words, the foreign owned enterprises impose relatively deprived productivity growth within the local firms. The reasons are to be found in the impulsive domestic competition evoked by the foreign investors and the lack of financial possibility for local firms to adapt properly within the new circumstances. Consequently, competitiveness of the entire economy might be improved only by locking up the particular gap in order to make the growth and convergence sustainable. FDI inflows may possibly increase productivity in the European transition economies principally affecting their exports performances so as to alleviate the balance of payments constraints. Many transition economies have been plunged into the foreign payments problems principally due to the lack of FDI inflows. Consequently, the policy makers have more counted on this source of external financing. Yet, some recent findings indicate that FDI concentration into non-tradable sector might undermine the export competitiveness and get the country up to greater exposure on economic crises. At the outset, noteworthy is to say that policy makers have to consider more active measures aimed at taking full advantage of FDI inflows, especially those that might create backward and forward linkages. The possibility to implement such measures, however, is limited either for the international commitments or the domestic restraints. Taking this into consideration, the best way to attract investors is to perk up the business environment, predominantly by employing predictable polices, transparent legal system and simple licensing regime.

## REFERENCES

1. Aghion, P. and Howitt, P. — "Capital, Innovation and growth accounting", Oxford Review of Economic Policy, 23(1), pp. 79-93, 2007
2. Aitken, B. and Harrison, A.E. — "Do domestic firms benefit from foreign investment? Evidence from Venezuela, American Economic Review, 89, pp. 605-618, 1999
3. Borensztein, E., Gregorio, J. and Lee, J.W. — "How does foreign direct investment affect economic growth?" NBER Working Paper (5057, 1995
4. Coe, D. and Helpman, E. — "International R&D spillovers", European Economic Review, Vol. 39, pp. 859-887, 1995
5. Eichengreen, B. — "Capital flows and crises", The MIT Press, 2004
6. Hunya, G. — "International Competitiveness : Impacts of FDI in CEECs ", The Vienna Institute for International Economic Studies (WIIW), Research Reports, No. 268, Vienna, 2000
7. Nesvera, K. — "The development of investment in 1991 to 1994 and Prognosis for 1995", Investicni a Postovani Banka, Newsletter for foreign investors (2), 1995
8. Romer, P. — "Increasing returns and long-run growth", Journal of Political Economy, Vol. 94, pp. 1002-1037, 1986
9. Sinn, H.W. and Weichenrieder, A. — "FDI, Political Resentment and the Privatization Process in Eastern Europe, CEPR, CES, MSH pp. 177-210, 1997
10. Sohinger, J. — "Transforming competitiveness in European transition economies: the role of Foreign direct investment", University of California Berkeley: Institute of European Studies, Political Economy of International Finance. Working Paper PEIF-17, May 2004
11. Stankovsky, J. — "Direct investment in Eastern Europe. Factors, Extent and Industry Structure", Study by WIFO commissioned by Bank Austria, Vienna, 1995
12. UNCTAD — "World Investment Reports" 1999, 2001, 2002, 2003, 2006